\documentclass[singlecolumn,aps,preprint]{revtex4-1}
\usepackage{amsfonts}
\usepackage{amssymb}
\usepackage{amsmath}
\usepackage{graphicx}
\usepackage{epsfig}
\usepackage{makecell}
\usepackage{enumerate}
\usepackage{mathrsfs}
\usepackage{color,xcolor}

\begin{document}

\title{Synchronized in-gap edge states and robust copropagation in
topological insulators without magnetic flux}
\author{Liangcai Xie}
\author{Tianyi He}
\author{Liang Jin}
\email{jinliang@nankai.edu.cn}
\affiliation{School of Physics, Nankai University, Tianjin 300071, China}

\begin{abstract}
\textbf{Abstract:} Copropagation of antichiral edge states in the metallic
phase requires the bulk states as counterpropagating modes. Without the band
gap protection, the copropagation along the boundaries is easily scattered
into the bulk and the counterpropagation in the bulk is not robust against
disorder and defects. Here, we propose a novel time-reversal symmetric
topological insulator holding the synchronized in-gap edge states. To
prevent the participation of bulk states, we introduce the time-reversal
symmetry to detach the edge states from the bulk band, then the robust
copropagation is realized through widening the band gap across a
metal-insulator transition with anisotropic next-nearest-neighbor couplings.
The time-reversal symmetry ensures the in-gap edge states with opposite
momenta as the counterpropagating modes. The inversion symmetry protects the
quantized polarization as a topological invariant. The synchronized in-gap
edge states not only enrich the family of topological edge states, but also
provide additional flexibility in the design of reconfigurable topological
optical devices. Our findings open a new avenue for the tailored robust wave
transport using the in-gap edge states for future acousto-optic topological
metamaterials.

\end{abstract}

\maketitle

\section*{1. Introduction}

Topological states in topological phases are robust against disorder and
defects \cite{FHPRL08}. In topological insulators, the chiral
edge states unidirectionally propagate in the opposite directions along the
parallel boundaries. The band gap protects the robustness of chiral edge
states. The helical edge states are constituted by a pair of time-reversal
symmetric chiral edge states associated with the pseudospin-up and
pseudospin-down \cite{MHNP11,ABKNM13,MCRNa13,ZHHangPRL18}. The valley is an
independent degree of freedom similar to the pseudospin~\cite%
{BZhangNP18,SZhangPRL20,XFRenPRL21}. The time-reversal symmetry of
each individual pseudospin is broken in the spin-Hall phase and the
inversion symmetry is broken in the valley-Hall phase \cite{LLu14}. The
propagations of helical edge states in spin-Hall phase or kink edge states
in valley-Hall phase are pinned to the pseudospins or valleys. The
helical/kink edge states associated with the same pseudospin/valley along
the parallel boundaries are counterpropagating modes for each other.

In topological metals, the antichiral edge states unidirectionally
copropagate in the same direction along the parallel boundaries \cite%
{FranzPRL18}. The band gap is absent and the on-resonant bulk states act as
counterpropagating channels for the antichiral edge states \cite%
{LiZYPRB20,ZhangBLPRL20}. Through reversing the magnetic flux associated
with one of the two sublattices, the topological phase changes from
insulator to metal and the chiral edge states become antichiral edge states
\cite{ZilberbergPRB20,ZLiuPRL23,LJinSciBull23}. Without the band
gap protection, once the antichiral excitation copropagating along the
boundaries is scattered into the bulk by disorder or at the corners, the
recovery of copropagation is impossible. An intriguing and important
question is whether it is possible to find copropagating edge states that
can avoid the participation of bulk states in the realization of
copropagation?

\begin{figure}[thb]
\includegraphics[bb=0 0 500 130,width=16 cm]{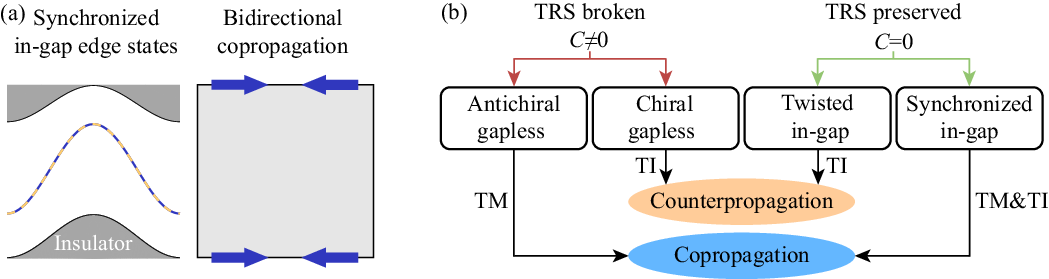}
\caption{(a) Schematics of the energy band structure for the synchronized
in-gap edge states and the associated bidirectional copropagation. (b)
Schematic of distinct types of two-dimensional edge states and their
propagations. TM stands for topological metal and TI stands for topological
insulator.} \label{fig1}
\end{figure}

Here, we find a novel time-reversal symmetric topological insulator that
hosts a pair of degenerate in-gap edge states, which synchronously
copropagate along the parallel boundaries back and forth against disorder
without being scattered into the bulk [Fig.~\ref{fig1}(a)]. This is realized
by widening the band gap until a metal-insulator phase transition. In the
topological insulating phase, the wide band gap provides a strong
topological protection, prohibits the counterpropagation in the bulk, and
protects the robustness in the copropagation. The time-reversal symmetry
ensures the bidirectionality. The inversion symmetry ensures the
synchronization. Furthermore, we demonstrate the disorder-immune robust
copropagation along the parallel boundaries for the edge state excitation.
The novel topological phase, supporting intriguing topological wave
transports mediated by the anisotropic long-range couplings, fundamentally
differs from the spin-Hall phase, valley-Hall phase, and high-order
topological phase. The flux-free setup without breaking the time-reversal
symmetry facilitates the experimental realization in topological
metamaterials.

The elaboration of our findings is organized as follows. We start by
introducing the definitions of gapless and in-gap edge states. Then, we
elucidate different types of two-dimensional edge states from the viewpoints
of time-reversal symmetry, band structure, and robust propagation dynamics
along the boundaries. The features of synchronized in-gap edge states are
revealed and highlighted from the demonstrations of time-reversal symmetry
breaking/preserving robust counterpropagations/copropagations exemplified in
the extended Qi-Wu-Zhang model, which is a square lattice having nontrivial
topology only in one direction. Furthermore, we propose a time-reversal
symmetric square lattice without magnetic flux holding the topological
insulating phase with synchronized in-gap edge states in both directions by
using the unbalanced reciprocal next-nearest-neighbor couplings. Finally, we
perform numerical simulations to show the robust copropagation of
synchronized in-gap edge states along the parallel boundaries in the
presence of random disorder.

\section*{2. Results and discussion}

Topological edge states are defined as gapless and in-gap from relative
position between the edge states $E(\mathbf{k)}$ and bulk bands $\varepsilon
_{\pm }(\mathbf{k})$. The edge states are called gapless if $E(\mathbf{k)}%
=\varepsilon _{\pm }(\mathbf{k})$ for certain $\mathbf{k}$ in the Brillouin
zone (BZ); i.e., the edge states are contacted with two adjacent bulk bands.
The edge states are called in-gap if $E(\mathbf{k)}\neq \varepsilon _{\pm }(%
\mathbf{k})$ for all $\mathbf{k}\in \mathrm{BZ}$; i.e., the edge states are
detached from the bulk band. The in-gap edge states spread in the entire BZ.
Consequently, the in-gap edge states support \textit{bidirectional}
propagation unless they are fully flat. In particular, the edge states are
called fully in-gap if $E(\mathbf{k)}\neq \varepsilon _{\pm }(\mathbf{k}%
^{\prime })$ for all $\mathbf{k}$, $\mathbf{k}^{\prime }\in \mathrm{BZ}$;
i.e., the edge states fully reside within the band gap. The gapless and
in-gap edge states may exist in the insulators and metals, whereas the fully
in-gap edge states only exist in the insulators.

\begin{figure}[t]
\includegraphics[bb=0 0 580 190,width=17.2cm]{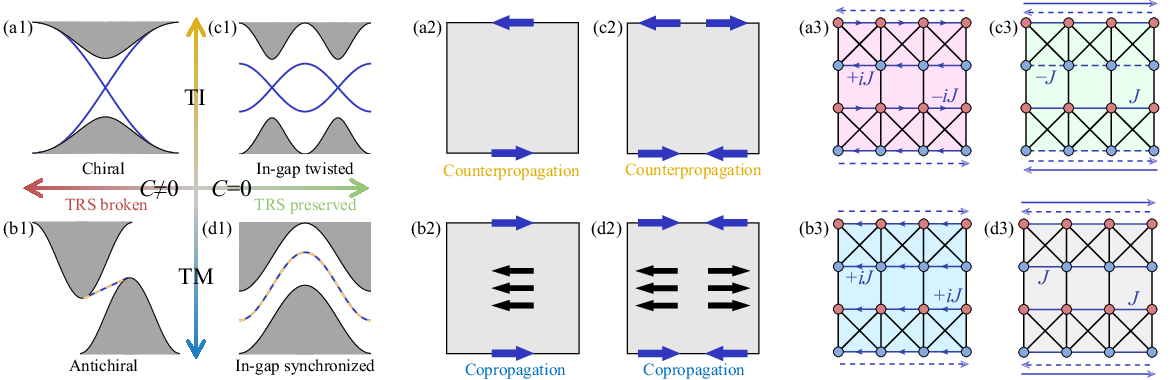}
\caption{Schematics of robust propagations of edge states along the parallel
boundaries, the associated band structures, and their realizations in the square lattices. (a) Unidirectional
counterpropagation of chiral edge states. (b) Unidirectional copropagation of  antichiral edge states. (c) Bidirectional
counterpropagation of twisted in-gap edge states. (d) Bidirectional copropagation of  synchronized in-gap edge states. In (a) and (b), the
time-reversal symmetry breaks and the edge states are gapless. In (c) and
(d), the time-reversal symmetry preserves and the edge states are in-gap.
The inversion symmetry results in the degeneracy of the in-gap edge states
in (d). Copropagation in the insulators is realized after the
metal-insulator transition by widening the band gap in (d) as illustrated
in Fig.~\protect\ref{fig1}(a).} \label{fig2}
\end{figure}

Figure~\ref{fig1}(b) concisely summarizes the two-dimensional edge states in
the first-order topological phases from the viewpoints of the time-reversal
symmetry breaking or preserving, the gapless or in-gap, and the metallic
phase or insulating phase. Figure~\ref{fig2} provides the schematics of the
band structures, the ways of propagations, and the concrete lattice
realizations of different types of edge states. More specifically, the
chiral (antichiral) edge states are the gapless edge states associated with
the topological insulators (metals) as elucidated in Fig.~\ref{fig2}(a1)
[Fig.~\ref{fig2}(b1)], and the unidirectional counterpropagation
(copropagation) along the parallel boundaries is schematically illustrated
in Fig.~\ref{fig2}(a2) [Fig.~\ref{fig2}(b2)] as a consequence of
time-reversal symmetry breaking. For the gapless edge states, the
metal-insulator phase transition alters the energy band from a topological
insulator to a topological metal and the topological edge states alter from
chiral to antichiral. This fundamentally stems from the fact that the edge
states are gapless, i.e., the bulk bands and the edge states are touched.
This causes the bulk states to participate as the counterpropagating modes
of antichiral edge states in a closed loop of propagation dynamics.

To prevent the participation of bulk states, we introduce the time-reversal
symmetry to add a backward going channel at the edges. The antichiral edge
states are altered into the in-gap edge states, i.e., the bulk band and\ the
edge states are detached from each other. We emphasize that although the
time-reversal symmetry ensures a zero Chern number, the topological phase
might still be nontrivial. Time-reversal symmetric insulators with
nontrivial topology may support the twisted in-gap edge states [Fig.~\ref%
{fig2}(c1)]. Interestingly, we discover novel time-reversal symmetric
topological insulators supporting the synchronized\ in-gap edge states [Fig.~%
\ref{fig2}(d1)]. The time-reversal symmetry guarantees the spectrum at\ the
opposite momenta having the same energy and opposite velocities. Thus, the
in-gap edge states include both the forward-going and backward-going edge
modes, and the boundaries support bidirectional light flow. In Fig.~\ref%
{fig2}(c2) [Fig.~\ref{fig2}(d2)], the edge state excitations with the same
momentum in the time-reversal symmetric insulators (metals) counterpropagate
(copropagate) along the parallel boundaries in the opposite (same)
directions with the same velocity.

In time-reversal symmetric topological phases, we emphasize that the
copropagation may appear in both the metallic [Fig.~\ref{fig2}(d1)] and
insulating phases [Fig.~\ref{fig1}(a)]. For the copropagation of in-gap edge
states in the metallic phase, the edge and bulk states both participate as
counterpropagating modes [Fig.~\ref{fig2}(d1)]. Widening the energy bands
illustrated in Fig.~\ref{fig2}(d1) alters the metallic phase into the
insulating phase, meanwhile the in-gap edge states are altered into the
fully in-gap edge states illustrated in Fig.~\ref{fig1}(a). For the
copropagation of fully in-gap edge states in the insulating phase, only the
edge states participate as the counterpropagating modes. The band gap
protects that the edge and bulk states are off-resonant, and the
copropagation of fully in-gap edge states remains robust without being
scattered into the bulk. The excitation of in-gap edge states on the
parallel boundaries with the same momenta realizes the robust copropagation
in the same directions as protected by the inversion symmetry.

Now, we consider the concrete lattices holding the topological phases and
edge states discussed previously in this section. The Haldane model is a
honeycomb lattice with nonreciprocal next-nearest-neighbor couplings, which
breaks the time-reversal symmetry. The standard Haldane model with identical
magnetic flux associated with its two sublattices holds the topological
insulating phase \cite{FHPRL08}. The modified Haldane model with opposite
magnetic fluxes associated with its two sublattices holds the topological
metallic phase \cite{FranzPRL18}. The nonzero Chern numbers predict the
chiral and antichiral edge states in both cases. To generate a time-reversal
symmetric topological phase, a simple consideration is the removal of
magnetic fluxes in the Haldane model. The standard Haldane model without
magnetic flux is a hexagonal lattice with reciprocal next-nearest-neighbor
couplings. Unfortunately, such a time-reversal symmetric hexagonal lattice
\textit{cannot} be tuned into an insulator due to the inseparable energy
bands. To propose the time-reversal symmetric topological phases with
separable energy bands with a band gap and the in-gap edge states, we
alternatively consider the extended Qi-Wu-Zhang model, which is a square
lattice with the reciprocal next-nearest-neighbor couplings uniformly
presented in one direction and alternatively presented in the other
direction \cite{LeykamPRL18}.

The time-reversal symmetry breaks in both the square lattices shown in Fig.~%
\ref{fig2}(a3) and Fig.~\ref{fig2}(b3). Figure~\ref{fig2}(a3) is the square
lattice having the insulating phase with the band structure shown in Fig.~%
\ref{fig2}(a1) and supporting the chiral edge states. In this configuration,
the nonreciprocal couplings along the top and bottom boundaries are opposite
and the magnetic fluxes associated with the upper and lower triangles are
identical. Figure~\ref{fig2}(b3) is the square lattice having the metallic
phase with the energy band structure shown in Fig.~\ref{fig2}(b1) and
supporting the antichiral edge states. In this configuration, the
nonreciprocal couplings along the top and bottom boundaries are identical
and the magnetic fluxes associated with the upper and lower triangles are
opposite. The time-reversal symmetry is preserved in both the square
lattices shown in Fig.~\ref{fig2}(c3) and Fig.~\ref{fig2}(d3). Figure~\ref%
{fig2}(c3) is the square lattice having the insulating phase with the band
structure shown in Fig.~\ref{fig2}(c1) and supporting the twisted in-gap
edge states. In this configuration, the reciprocal couplings along the top
and bottom boundaries have opposite signs. Figure~\ref{fig2}(d3) is the
square lattice having the metallic phase with the band structure shown in
Fig.~\ref{fig2}(d1) and supporting the synchronized in-gap edge states. In
this configuration, the reciprocal couplings along the top and bottom
boundaries have identical sign. The synchronized in-gap edge states in Fig.~%
\ref{fig2}(d1) are protected by the inversion symmetry. The metallic phase
alters to the insulating phase via widening the band gap; then, the
topological insulators holding robust copropagation of in-gap edge states
without being scattered into the bulk are realized.

The time-reversal symmetric topological phases have a zero Chern number, but
there exists another topological invariant to characterize the band topology
\cite{LJinPRL24}. Notably, the time-reversal symmetry and inversion symmetry
jointly protect the vanishing of Berry curvature in the entire BZ, and the
Berry connection does not have any singularity. In this situation, the
two-dimensional Zak phase, known as the polarization, is valid for
topological characterization
\begin{equation}
\mathbf{P}=(2\pi )^{-2}\int_{\mathrm{BZ}}\mathbf{A}(k_{x},k_{y})dk_{x}dk_{y},
\end{equation}%
where $\mathbf{A}(k_{x},k_{y})=-\langle \psi \left( \mathbf{k}\right)
|i\nabla _{\mathbf{k}}|\psi \left( \mathbf{k}\right) \rangle $ is the Berry
connection and $\left\vert \psi (\mathbf{k})\right\rangle $ is the Bloch
wave function. The two components of polarization $\mathbf{P}=\left(
P_{x},P_{y}\right) $ describe the topologies of the one-dimensional
projection lattices. $P_{x}\neq 0$ and $P_{y}\neq 0$ indicate the existence
of in-gap edge states under the open boundary condition (OBC) in the $x$-
and $y$-direction, respectively.

Although the extended Qi-Wu-Zhang model in the configuration Fig.~\ref{fig2}%
(d3) realizes a topological insulating phase holding the synchronized in-gap
edge states on the top and bottom boundaries, the band topology on the other
direction is trivial, and the edge states are absent on the left and right
boundaries. A topological insulating phase supporting the robust
copropagation in both directions greatly benefits the robust light steering
in topological metamaterials. This motivates us to further search for a
configuration with more flexible band topology in both directions.

\begin{figure}[tb]
\includegraphics[bb=0 0 570 175, width=17cm]{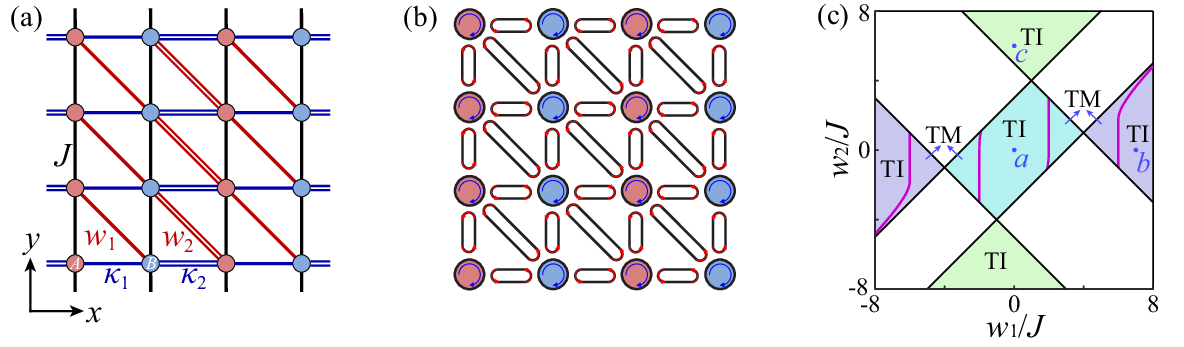}
\caption{(a) Schematic of the two-dimensional time-reversal symmetric square
lattice. (b) Schematic of the coupled-resonator realization. The ring
resonators in red (blue) are the sites $A$ ($B$), and the other stadium
resonators are the linking resonators. The blue (red) arrows in the
resonators indicate the clockwise (counterclockwise) mode. (c) Phase diagram
for $\protect\kappa _{1}=J,\protect\kappa _{2}=4J$.} \label{fig3}
\end{figure}

Based on the configuration in Fig.~\ref{fig2}(d3) by destroying the balanced
next-nearest-neighbor couplings and maintaining the staggered feature in one
direction, we propose a time-reversal symmetric square lattice without
magnetic flux as shown in Fig.~\ref{fig3}(a). Notably, the configuration in
Fig.~\ref{fig3}(a) also has the $C_{2}$ symmetry and is easy to realize in
optical waveguides \cite{MCRNa13}, resonators \cite{MHafeziNP16,YHYangNa19},
and photonic crystals \cite{BZhenNature20,HChenNC23}. The flourishing
development of topological photonics has spurred experimental simulations of
topological systems, providing researchers with platforms to explore novel
photonic topological phenomena. Resonators are widely used to simulate
topological systems due to their mode stability, high quality factor and
integrability. The nonreciprocal couplings are absent in the time-reversal
symmetric two-dimensional square lattice. Then, the coupled waveguide
arrays, acoustic, electric, and mechanical systems are suitable platforms
for the investigation of time-reversal symmetric topological phases without
any magnetic flux. In these physical systems, the couplings among the
lattice sites are created via directly linking the sites. Thus, the
next-nearest-neighbor couplings are unrelated to the nearest-neighbor
couplings in the square lattice, and all the couplings are independently
tunable.

Figure~\ref{fig3}(b) is the schematic of the coupled resonator realization
of the time-reversal symmetric two-dimensional square lattice. The ring
resonators are the primary resonators for the sites of the square lattice.
The primary resonators in red are the sites $A$ and primary resonators in
blue are the sites $B$. Each resonator supports both the clockwise and
counterclockwise modes and the intracavity modes are decoupled. The linking
resonators mediate the photons tunneling in the primary resonators and
induce the effective couplings between the nearest-neighbor primary
resonators and the next-nearest-neighbor primary resonators. The linking
resonators and the primary resonators are coupled through their evanescent
fields and the coupling strengths depend on the distance between the linking
resonators and the primary resonators \cite{MHNP11}. The time-reversal
symmetric square lattice only has reciprocal couplings. The coupling
strength $\kappa $ is approximately characterized by $\kappa =\kappa
_{l}^{2}/\Delta _{l}$, where $\kappa _{l}$ and $\Delta _{l}=\omega
_{c}-\omega _{link}$ are the hopping and detuning between the frequency of
primary resonators $\omega _{c}$ and the frequency of linking resonators $%
\omega _{link}$.

Applying the Fourier transformation to the Hamiltonian of the square lattice
in the real space, we obtain the Bloch Hamiltonian in the momentum space in
the form of
\begin{equation}
h\left( \mathbf{k}\right) =\mathbf{d\left( \mathbf{k}\right) \cdot \sigma +}%
d_{0}\mathbf{\left( \mathbf{k}\right) }\sigma _{0},
\end{equation}%
where $\mathbf{\sigma =}\left( \sigma _{x},\sigma _{y},\sigma _{z}\right) $
is the Pauli matrix and $\sigma _{0}$ is the identity matrix. The effective
magnetic field is $\mathbf{d\left( \mathbf{k}\right) =}[d_{x}\left( \mathbf{k%
}\right) ,d_{y}\left( \mathbf{k}\right) ,d_{z}\left( \mathbf{k}\right) ]$
with
\begin{eqnarray}
d_{x}\left( \mathbf{k}\right) &=&\kappa _{1}+\kappa _{2}\cos k_{x}+w_{1}\cos
k_{y}+w_{2}\cos \left( k_{x}-k_{y}\right) , \\
d_{y}\left( \mathbf{k}\right) &=&\kappa _{2}\sin k_{x}+w_{1}\sin
k_{y}+w_{2}\sin \left( k_{x}-k_{y}\right) , \\
d_{z}\left( \mathbf{k}\right) &=&0.
\end{eqnarray}%
The term $d_{0}\mathbf{\left( \mathbf{k}\right) =}2J\cos k_{y}$ varies the
energies of the Bloch bands
\begin{equation}
\varepsilon _{\pm }\left( \mathbf{k}\right) =d_{0}\mathbf{\left( \mathbf{k}%
\right) \pm \lbrack }d_{x}^{2}\mathbf{\left( \mathbf{k}\right) +}d_{y}^{2}%
\mathbf{\left( \mathbf{k}\right) ]}^{1/2},
\end{equation}%
however, the eigenstates of the Bloch bands are unaffected.

The polarization $\mathbf{P}=\left( P_{x},P_{y}\right) $ is associated with
the vector field $[d_{x}\left( \mathbf{k}\right) ,d_{y}\left( \mathbf{k}%
\right) ]$ (see Supplementary A). The nontrivial topology $\mathbf{P}%
\neq\left( 0,0\right) $ predicts the edge states in the separable region of
the projection spectra (see Supplementary B). The uniform nearest-neighbor
coupling $J$ in the vertical direction can open a band gap without
destroying the topology. The staggered nearest-neighbor couplings $\kappa
_{1}$, $\kappa _{2}$ in the horizontal direction and the anisotropic
next-nearest-neighbor couplings $w_{1}$, $w_{2}$ create the nontrivial
topology, adjust the edge states, and alter the robust dynamics. We
highlight the flexibility of nontrivial topology induced by the anisotropic
next-nearest-neighbor couplings. The copropagations in the horizontal and
vertical directions can be independently tunable. Topologically nontrivial
region in the $x$-direction characterized by $P_{x}=1/2$ satisfies $|\kappa
_{1}+w_{1}e^{ik_{y}}|<|\kappa _{2}+w_{2}e^{ik_{y}}|$, which predicts the
existence of edge states under the OBC in the $x$-direction. Topologically
nontrivial region in the $y$-direction characterized by $P_{y}=\pm 1/2$
satisfies $|\kappa _{1}+\kappa _{2}e^{ik_{x}}|<|w_{1}+w_{2}e^{ik_{x}}|$,
which predicts the existence of edge states under the OBC in the $y$%
-direction. The strong coupling $\kappa _{1}$ ($w_{2}$) destroys
(constructs) the topologies in both directions. The strong coupling $\kappa
_{2}$ ($w_{1}$) constructs (destroys) the topology in the $x$-direction and
destroys (constructs) the topology in the $y$-direction. The topological
phase transition occurs when the band gap closes at $w_{1}-w_{2}=\pm (\kappa
_{1}-\kappa _{2})$ and $w_{1}+w_{2}=\pm (\kappa _{1}+\kappa _{2})$.

\begin{figure}[tb]
{\centering\includegraphics[bb=0 0 261 260, width=15cm]{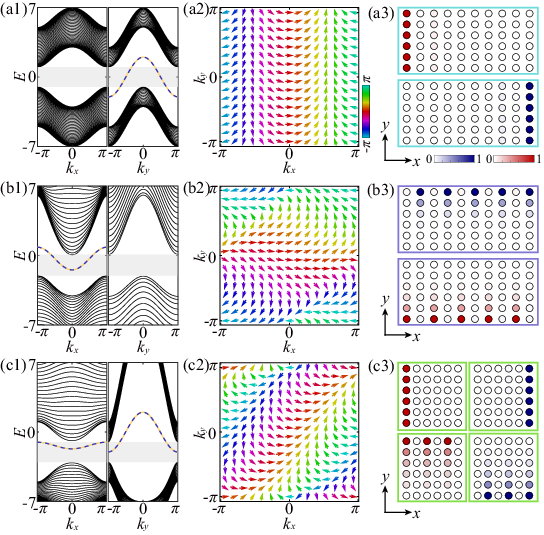}}
\caption{The projection energy band structure (left panel), the visualized
polarization (middle panel), and the edge state distribution (right panel)
for (a) $w_{1}=0,w_{2}=0$, (b) $w_{1}=7J,w_{2}=0$, and (c) $w_{1}=0,w_{2}=6J$
as marked in Fig.~\protect\ref{fig3}(c). The angles of the arrows represent $\arctan [d_{y}\left( \mathbf{k}\right) /d_{x}\left( \mathbf{k}\right) ]$ at
the momentum $\mathbf{k}=(k_{x},k_{y})$. The values of polarization are (a2) $P_{x}=1/2$, $P_{y}=0$, (b2) $P_{x}=0$, $P_{y}=1/2$, and (c2) $P_{x}=1/2$, $P_{y}=-1/2$, respectively.}
\label{fig4}
\end{figure}

The profile of phase diagram in the parameter space $w_{1}$-$w_{2}$ [Fig.~%
\ref{fig3}(c)] depends on the ratio of $\kappa _{1}$\ and $\kappa _{2}$. The
colored regions are gapped phases. The band gap ensures the separation of
the bulk and edge states. The magenta lines indicate the metal-insulator
phase transition. In the transition from a topological metal to a
topological insulator, the band gap widens without a topological phase
transition; and the edge states change from in-gap to fully in-gap, allowing
the copropagation without the participation of bulk states as the
counterpropagating modes. $w_{1}$, $w_{2}$, and\ $\kappa _{2}$\ increase the
band gap to enter topological insulating phase. The white regions are
gapless phase with band touching.

The polarization in the cyan region is $(P_{x},P_{y})=(1/2,0)$ for $\kappa
_{1}<\kappa _{2}$. A pair of edge states appear under the OBC in the $x$
direction [Fig.~\ref{fig4}(a1)]. The arrow at certain $k_{y}$
counterclockwise rotates once as $k_{x}$ varies an entire period from $-\pi $
to $\pi $ [Fig.~\ref{fig4}(a2)], which corresponds to $P_{x}=1/2$. The edge
states $E_{L}$ and $E_{R}$ appear on the left and right boundaries [Fig.~\ref%
{fig4}(a3)] and copropagate along the vertical direction. The polarization
in the purple regions is $(P_{x},P_{y})=(0,1/2)$, a pair of edge states
appear under the OBC in the $y$ direction [Fig.~\ref{fig4}(b1)]. The arrow
at certain $k_{x}$ counterclockwise rotates once as $k_{y}$ varies an entire
period from $-\pi $ to $\pi $ [Fig.~\ref{fig4}(b2)], which corresponds to $%
P_{y}=1/2$. The edge states $E_{T}$ and $E_{B}$ appear on the top and bottom
boundaries [Fig.~\ref{fig4}(b3)] and copropagate along the horizontal
direction. As illustrated in Fig.~\ref{fig4}(a3) and Fig.~\ref{fig4}(b3),
the edge states localized on the left and bottom (right and top) boundaries
are mostly localized on the sublattice $A$ ($B$). The polarization in the
green regions is $(P_{x},P_{y})=(1/2,-1/2) $. A pair of edge states appear
under the OBC in the $x$- and $y$-direction, respectively [Fig.~\ref{fig4}%
(c1)]. The arrow at certain $k_{y}$ ($k_{x}$) counterclockwise (clockwise)
rotates once as $k_{x} $ ($k_{y}$) varies an entire period from $-\pi $ to $%
\pi $ [Fig.~\ref{fig4}(c2)], which corresponds to $P_{x}=1/2$ ($P_{y}=-1/2$%
). The edge state localized on the left (right) boundary is still localized
on the sublattice $A$ ($B$), but the edge state localized on the bottom
(top) boundary changes to be mostly localized on the sublattice $B$ ($A$)
[Fig.~\ref{fig4}(c3)]. Topological phases with $P_{y}=1/2$ satisfy $%
\left\vert w_{1}\right\vert >\left\vert w_{2}\right\vert $ and topological
phases with $P_{y}=-1/2$\ satisfy $\left\vert w_{1}\right\vert <\left\vert
w_{2}\right\vert $, then, the edge states $E_{T}$ and $E_{B}$ mainly occupy
different sublattices (see Supplementary C). The square lattice lacks $C_{4}$
symmetry protection, thus $P_{x}\cdot P_{y}\neq 0$ is a first-order
topological phase~\cite{HughesPRB19}. The edge states present on the four
boundaries and copropagate along both the vertical and horizontal
directions. Although the four boundaries are bidirectional, the edge states
\textit{cannot} clockwise or counterclockwise circulate along the boundaries
similar to two pairs of opposite chiral edge states. In addition, $%
P_{x}\cdot P_{y}\neq 0$ ($=0$) does not indicate the presence (absence) of
corner states \cite{HughesScience17,FLiuPRL19,JGongPRL19} (see Supplementary
D).

Notably, the robust propagation dynamics associated with the edge states are
important in practical applications. The robustness to obstacles is a
consequence of the unidirectionality from the time-reversal symmetry
breaking, where the counterpropagating channel is closed and the
backscattering is completely suppressed. This is a typical feature of the
chiral edge states. In contrast to the unidirectional wave propagation of
chiral edge states, we emphasize that the bidirectional wave propagation of
the synchronized in-gap edge states cannot robustly pass through the
intentionally embedded obstacles due to the presence of backscattering
channel as a consequence of the time-reversal symmetry. Nevertheless, the
propagation dynamics of the in-gap edge states are still robust to random
disorder. The robustness originates from the band gap protection. In
practice, the use of edge states for topological transport seldom encounters
obstacles, and the propagations along the boundaries are rarely affected by
the random fabrication errors and environmental perturbations without the
emergence of obvious backscattering. Thus, the robustness of in-gap edge
states protected by the wide band gap is enough for the practical
applications. The robustness of edge states to certain disorder closely
depends on whether the disorder breaks the symmetry of the topological
system. The degeneracy of in-gap edge states is destroyed when the inversion
symmetry is absent in the presence of random disorder. Although the edge
states are slightly affected by the disorder with a splitting in the energy,
the band gap is still able to provide the topological protection for their
propagation dynamics.

\begin{figure*}[th]
\includegraphics[bb=0 0 420 360, width=14.0cm]{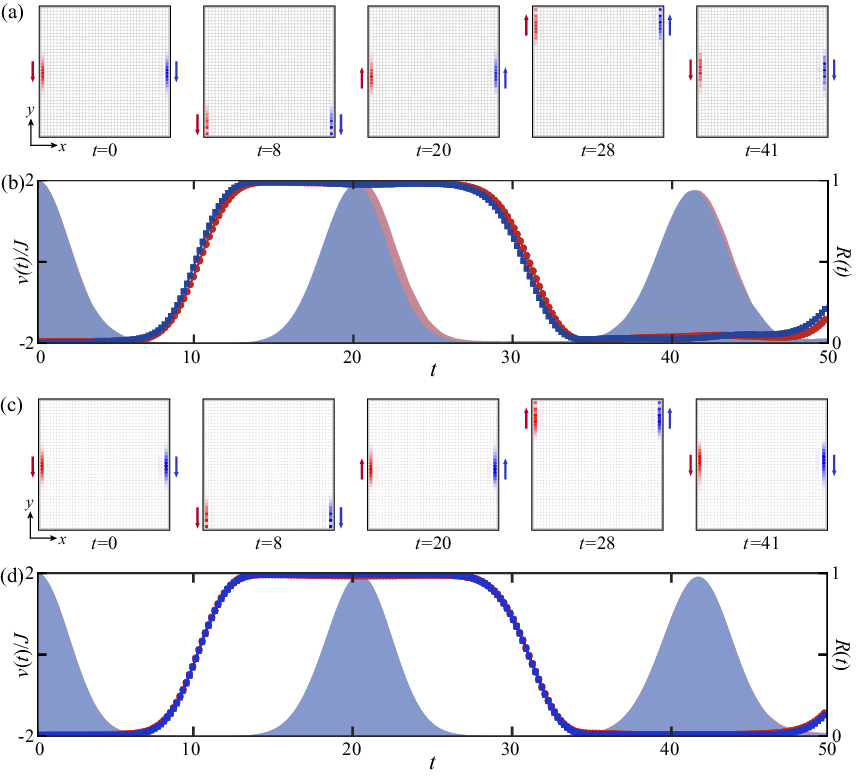}
\caption{(a) Snapshots of copropagation along the vertical boundaries and
(b) the propagation velocity and return probability for the couplings
randomly deviated from the set strengths within the range of $[-5\%,5\%]$.
(c) Snapshots of copropagation along the vertical boundaries and (d) the
propagation velocity and return probability for the detunings randomly
deviated within the range of $[-5\%,5\%]$ of the unit coupling $J$. The
initial excitation is the Gaussian wave packet of the edge state with the
momentum $k_{y}=\protect\pi/2$. Red and blue lines indicate the propagation
velocities $\protect\nu (t)$ of the wave packets on the left and right
boundaries, respectively. Red and blue areas indicate the return
probabilities $R(t)$ of the wave packets on the left and right boundaries,
respectively. The lattice size is $40\times 40$, the system parameters are
from Fig.~\protect\ref{fig4}(a). The unit of time is $J^{-1}$.}
\label{fig5}
\end{figure*}

To verify our theoretical analysis, we use the Gaussian wave packet to simulate the propagation dynamics \cite{LJinSciBull23, LJinPRL24, JHJiangNC24, BoYan24, FangweiYe24}. Figure~\ref{fig5} numerically demonstrates the robust
copropagation of Gaussian wave packet of the edge state excitation from the
topological phase $(P_{x},P_{y})=(1/2,0)$ in the presence of random
disorder. The left (right) edge state $E_{L}$ ($E_{R}$) is exactly localized
on the sublattice $A$ ($B$). The edge state energy is $E_{L,R}\left(
k_{y}\right) =2J\cos k_{y}$. The velocity of edge state excitation is $\nu
_{L,R}\left( k_{y}\right) =dE_{L,R}\left( k_{y}\right) /dk_{y}$. In the
numerical simulations, the initial excitation $\left\vert \Psi
(0)\right\rangle $ is a Gaussian wave packet of the edge state localized on
the left/right boundary
\begin{equation}
\left\vert \Psi (0)\right\rangle =\Omega
\sum\nolimits_{k_{y}}e^{-(k_{y}-k_{0})^{2}/(2\alpha
^{2})}e^{-iN_{c}(k_{y}-k_{0})}e^{ik_{y}n}{\left\vert \psi
_{L/R}\right\rangle ,}
\end{equation}%
where $\left\vert \psi _{L/R}\right\rangle $ is the wave function of the
left/right edge state in the momentum space (see Supplementary C) and $n$ is
the index of unit cell. $k_{0}$ is the central momentum, $N_{c}$ is the
center, and $\Omega $ is the normalization coefficient of the Gaussian wave
packet. The parameter $\alpha $ controls the width of the Gaussian wave
packet. A Gaussian wave packet in momentum space under the
Fourier transformation remains a Gaussian wave packet in real space as shown in Fig.~\ref{fig5}. $\left\vert \Psi (t)\right\rangle $ is the time
evolution of the initial excitation $\left\vert \Psi (0)\right\rangle $ in
the square lattice under OBCs on both directions at the moment $t$. The
return probability is $R(t)=\sum_{j}\left\vert \Psi _{j}(0)\Psi
_{j}(t)\right\vert ^{2}$, where the subscript $j$ is the site number. The dynamics of the Gaussian wave packet
excitation reflects the dispersion of the edge state at the momentum $k_{0}$ in the band structure under open boundary condition of the square lattice as shown in Fig.~\ref{fig4}%
. The square lattice size is $40\times 40$, the Gaussian wave packet returns to its initial position with an opposite velocity at about the
moment $t=20J^{-1}$ for the first time. Thus, the velocity of the edge state excitation along the boundary in
the numerical simulation is about $2J$. This is consistent with the propagation velocity $\nu_{L,R}\left( k_{0}\right)$ obtained in our theoretical analysis. In addition, the edge state excitations with opposite momenta are degenerate, and they propagate in opposite directions. However, the degenerate edge state excitations in their propagations do not merge with each other or backward propagate until being reflected back at the corner of the square lattice.

The edge state excitations on both the left and right boundaries at the
central momentum $k_{y}=\pi /2$ in the presence of random coupling disorder
within the range of $[-5\%,5\%]$ are performed in Figs.~\ref{fig5}(a) and~%
\ref{fig5}(b). Figure~\ref{fig5}(a) provides the snapshots for the
synchronized copropagation. Figure~\ref{fig5}(b) exhibits the velocity and
return probability during the copropagation. The coupling disorder does not
obviously affect the robust copropagation of the edge states even if the
deviation from the exact values of couplings reaches the range of $[-10\%,10\%]$ (see Supplementary E). The random detuning in the resonator
frequency is the on-site Anderson disorder of the time-reversal symmetric
square lattice. Although the Anderson disorder may solely create nontrivial
band topology known as the topological Anderson insulator \cite%
{SQShenPRL09,BLZhangPRL20,JWDongPRL24}, the random detunings in the
time-reversal symmetric square lattice break the inversion symmetry of the
square lattice and slightly disturb the edge states. Nevertheless, the
copropagation is robust to detuning disorder under the band gap protection.
The edge state excitations on both the left and right boundaries at the
central momentum $k_{y}=\pi /2$ in the presence of random detuning disorder
are performed in Figs.~\ref{fig5}(c) and~\ref{fig5}(d). Figure~\ref{fig5}(c)
provides the snapshots for the synchronized copropagation. Figure~\ref{fig5}%
(d) exhibits the velocity and return probability during the copropagation.
In the numerical simulations, the Gaussian wave packets copropagate along
the boundaries, and bounce back and forth at the corners without obvious
backscattering in the copropagation for weak disorder. The Gaussian wave
packets complete a cyclic process after returning to the initial position
with the same velocity. The disorder-immune robust copropagation dynamics of
the edge state excitation verifies the theoretical prediction on the
robustness.

In addition, the bidirectionality of the in-gap edge states protected by the
time-reversal symmetry naturally provides a possibility for picking the
propagation direction along the boundaries in a desirable manner via selectively choosing the momenta of edge state excitations. This
offers a topological insulator to simultaneously support robust
copropagation and counterpropagation. The disorder-immune robust
counterpropagation is realized by selectively exciting the in-gap edge
states with the momentum $k_{y}=\pi /2$ on the left boundary and the
momentum $k_{y}=-\pi /2$ on the right boundary (see Supplementary F).

\section*{3. Conclusions}

In conclusion, we have proposed a novel time-reversal symmetric topological
insulator supporting the synchronized in-gap edge states. The associated
copropagation dynamics avoids the participation of bulk states and prevents
the edge state excitations from being scattered into the bulk. The discovery
of synchronized in-gap edge states in the time-reversal symmetric
topological insulators has expanded our understanding of two-dimensional
topological phases of matter, enriched the band topology, and provided new
opportunities for topological light steering. The robust copropagation can
be generalized to higher-dimension \cite{JiaPRB21,JWDongNC23,ZGaoNC23} and
higher-order topology \cite{Rho20}. It is interesting to further explore
robust copropagation in nonlinear \cite{ZChenScience21} and non-Hermitian
\cite{YAPRL20,YHLPRL22,ZYangPRL24} photonic systems. In addition, the robust
copropagation along the parallel boundaries provides versatile ways of wave
transport for robust light steering \cite{HPJPP22}. The unique topology in
the time-reversal symmetric square lattice enriched by the anisotropic
long-range couplings is highlighted. It is also convenient to realize
topological lattices without magnetic flux in the acoustic \cite%
{MXNP15,YFChenNP16,YGPeng16,ZLiuPRL17,ZLiuPRL18,JiangNP19,LLuPRL20,CQiuPRL22}%
, electronic \cite{XDZhangPRL21}, and mechanical \cite{GMaNRP19} systems.
Our findings pave the way for future explorations on the fundamentals and
applications of time-reversal symmetric topological metamaterials.

\section*{Declaration of competing interest}

The authors declare that they have no conflicts of interest in this work.

\section*{Acknowledgments}

This work was supported by National Natural Science Foundation of China
(Grant No.~12222504).

\section*{Supplementary materials}

Supplementary A provides the polarization for topological characterization.
Supplementary B provides the edge states in the topological metallic phase.
Supplementary C provides the energies and wave functions of edge states.
Supplementary D provides the difference between the proposed square lattice
and the two-dimensional Su-Schrieffer-Heeger model. Supplementary E provides
the robust copropagation of in-gap edge states in the presence of random
coupling disorder within the range of $[-10\%,10\%]$. Supplementary F
provides the robust counterpropagation of in-gap edge states for the
excitation on the parallel boundaries with the opposite momenta.


\begin{thebibliography}{99}
\bibitem{FHPRL08} F. D. M. Haldane, S. Raghu, Possible Realization of
Directional Optical Waveguides in Photonic Crystals with Broken
Time-Reversal Symmetry, Phys. Rev. Lett. 100 (2008) 013904.

\bibitem{MHNP11} M. Hafezi, E. A. Demler, M. D. Lukin, et al., Robust
optical delay lines with topological protection, Nat. Phys. 7 (2011) 907.

\bibitem{ABKNM13} A. B. Khanikaev, S. H. Mousavi, W.-K. Tse, et al.,
Photonic topological insulators, Nat. Mater. 12 (2013) 233.

\bibitem{MCRNa13} M. C. Rechtsman, J. M. Zeuner, Y. Plotnik, et al.,
Photonic Floquet Topological Insulators, Nature 496 (2013) 196.

\bibitem{ZHHangPRL18} Y. Yang, Y. F. Xu, T. Xu, et al., Visualization of
unidirectional optical waveguide using topological photonic crystals made of
dielectric material, Phys. Rev. Lett. 120 (2018) 217401.

\bibitem{BZhangNP18} F. Gao, H. Xue, Z. Yang, et al., Topologically
protected refraction of robust kink states in valley photonic crystals, Nat.
Phys. 14 (2018) 140.

\bibitem{SZhangPRL20} H. F. Yang, J. Xu, Z. F. Xiong, et al., Optically
Reconfigurable Spin-Valley Hall Effect of Light in Coupled Nonlinear Ring
Resonator Lattice, Phys. Rev. Lett. 127 (2020) 043904.

\bibitem{XFRenPRL21} Y. Chen, X. T. He, Y. J. Cheng, et al., Topologically
Protected Valley-Dependent Quantum Photonic Circuits, Phys. Rev. Lett. 126
(2021) 230503.

\bibitem{LLu14} L. Lu, J. D. Joannopoulos, M. Solja\v{c}i\'{c}, Topological
photonics, Nat. Photon. 8 (2014) 821.

\bibitem{FranzPRL18} E. Colom\'{e}s, M. Franz, Antichiral Edge States in a
Modified Haldane Nanoribbon, Phys. Rev. Lett. 120 (2018) 086603.

\bibitem{LiZYPRB20} J. Chen, W. Liang, Z. Y. Li, Antichiral one-way edge
states in a gyromagnetic photonic crystal, Phys. Rev. B 101 (2020) 214102.

\bibitem{ZhangBLPRL20} P. Zhou, G. G. Liu, Y. Yang, et al., Observation of
Photonic Antichiral Edge States, Phys. Rev. Lett. 125 (2020) 263603.

\bibitem{ZilberbergPRB20} M. M. Denner, J. L. Lado, O. Zilberberg,
Antichiral states in twisted graphene multilayers, Phys. Rev. Research 2
(2020) 043190.

\bibitem{ZLiuPRL23} M. Yan, X. Huang, J. Wu, et al., Antichirality Emergent
in Type-II Weyl Phononic Crystals, Phys. Rev. Lett. 130 (2023) 266304.

\bibitem{LJinSciBull23} L. Xie, L. Jin, Z. Song, Antihelical edge states in
two-dimensional photonic topological metals, Sci. Bull. 68 (2023) 255.

\bibitem{LeykamPRL18} D. Leykam, S. Mittal, M. Hafezi, et al.,
Reconfigurable Topological Phases in Next-Nearest-Neighbor Coupled Resonator
Lattices, Phys. Rev. Lett. 121 (2018) 023901.

\bibitem{LJinPRL24} H. C. Wu, H. S. Xu, L. C. Xie, et al., Edge State, Band
Topology, and Time Boundary Effect in the Fine-Grained Categorization of
Chern Insulators, Phys. Rev. Lett. 132 (2024) 013904.

\bibitem{MHafeziNP16} S. Mittal, S. Ganeshan, J. Fan, et al., Measurement of
topological invariants in a 2D photonic system, Nat. Photon. 10 (2016) 180.

\bibitem{YHYangNa19} Y. Yang, Z. Gao, H. Xue, et al., Realization of a
three-dimensional photonic topological insulator, Nature 565 (2019) 622.

\bibitem{BZhenNature20} X. Yin, J. Jin, M. Solja\v{c}i\'{c}, et al.,
Observation of topologically enabled unidirectional guided resonances,
Nature 580 (2020) 467.

\bibitem{HChenNC23} X. X. Wang, Z. Guo, J. Song, et al., Unique
Huygens-Fresnel electromagnetic transportation of chiral Dirac wavelet in
topological photonic crystal, Nat. Commun. 14 (2023) 3040.

\bibitem{HughesPRB19} W. A. Benalcazar, T. Li, T. L. Hughes, Quantization of
fractional corner charge in Cn-symmetric higher-order topological
crystalline insulators, Phys. Rev. B 99 (2019) 245151.

\bibitem{HughesScience17} W. A. Benalcazar, B. A. Bernevig, T. L. Hughes,
Quantized electric multipole insulators, Science 357 (2017) 61.

\bibitem{FLiuPRL19} F. Liu, H. Y. Deng, K. Wakabayashi, Helical Topological
Edge States in a Quadrupole Phase, Phys. Rev. Lett. 122 (2019) 086804.

\bibitem{JGongPRL19} C. H. Lee, L. Li, J. Gong, Hybrid higher-order
skin-topological modes in nonreciprocal systems, Phys. Rev. Lett. 123 (2019)
016805.

\bibitem{JHJiangNC24} Z. Li, L. W. Wang, X. L. Wang, Observation of dynamic non-Hermitian skin effects, Nat. Commun. 15 (2024) 6544.

\bibitem{BoYan24} Z. L. Dong, H. Li, T. Wan, Quantum time reflection and refraction of ultracold atoms, Nat. Photon. 18 (2024) 68.

\bibitem{FangweiYe24} N. Khan, P. Wang, Q. D. Fu, Observation of period-doubling Bloch oscillations, Phys. Rev. Lett. 132 (2024) 053801.

\bibitem{SQShenPRL09} J. Li, R. L. Chu, J. K. Jain, et al., Topological
anderson insulator, Phys. Rev. Lett. 102 (2009) 136806.

\bibitem{BLZhangPRL20} G. G. Liu, Y. H. Yang, X. Ren, et al., Topological
Anderson insulator in disordered photonic crystals, Phys. Rev. Lett. 125
(2020) 133603.

\bibitem{JWDongPRL24} X. D. Chen, Z. X. Gao, X. H. Cui, et al., Realization
of time-reversal invariant photonic topological Anderson insulators, Phys. Rev. Lett. 133 (2024) 133802.

\bibitem{JiaPRB21} X. Cheng, J. Chen, L. Zhang, et al., Antichiral edge
states and hinge states based on the Haldane model, Phys. Rev. B 104 (2021)
L081401.

\bibitem{JWDongNC23} J. W. Liu, F. L. Shi, K. Shen, et al., Antichiral
surface states in time-reversal-invariant photonic semimetals, Nat. Commun.
14 (2023) 2027.

\bibitem{ZGaoNC23} X. Xi, B. Yan, L. Yang, et al., Topological antichiral
surface states in a magnetic Weyl photonic crystal, Nat. Commun. 14 (2023)
1991.

\bibitem{Rho20} M. Kim, Z. Jacob, J. Rho, Recent advances in 2D, 3D and
higher-order topological photonics, Light Sci. Appl. 9 (2020) 130.

\bibitem{ZChenScience21} S. Xia, D. Kaltsas, D. Song, et al., Nonlinear
tuning of PT symmetry and non-Hermitian topological states, Science 372
(2021) 72.

\bibitem{YAPRL20} Y. Ao, X. Hu, Y. You, et al., Topological Phase Transition
in the Non-Hermitian Coupled Resonator Array, Phys. Rev. Lett. 125 (2020)
013902.

\bibitem{YHLPRL22} Y. H. Li, C. Liang, C. Y. Wang, et al., Gain-Loss-Induced
Hybrid Skin-Topological Effect, Phys. Rev. Lett. 128 (2022) 223903.

\bibitem{ZYangPRL24} Y. Sun, X. Hou, T. Wan, et al., Photonic Floquet
Skin-Topological Effect, Phys. Rev. Lett. 132 (2024) 063804.

\bibitem{HPJPP22} H. Price, Y. Chong, A. Khanikaev, et al., Roadmap on
topological photonics, J. Phys. Photonics 4 (2022) 032501.

\bibitem{MXNP15} M. Xiao, W. Chen, W. He, et al., Synthetic gauge flux and
Weyl points in acoustic systems, Nat. Phys. 11 (2015) 920.

\bibitem{YFChenNP16} C. He, X. Ni, H. Ge, et al., Acoustic topological
insulator and robust one-way sound transport, Nat. Phys. 12 (2016) 1124.

\bibitem{YGPeng16} Y. G. Peng, C. Z. Qin, D. G. Zhao, et al., Experimental
demonstration of anomalous Floquet topological insulator for sound, Nat.
Commun. 7 (2016) 13368.

\bibitem{ZLiuPRL17} J. Lu, C. Qiu, L. Ye, et al., Observation of topological
valley transport of sound in sonic crystals, Nat. Phys. 13 (2017) 369.

\bibitem{ZLiuPRL18} J. Lu, C. Qiu, W. Deng, et al., Valley Topological
Phases in Bilayer Sonic Crystals, Phys. Rev. Lett. 120 (2018) 116802.

\bibitem{JiangNP19} X. Zhang, H. X. Wang, Z. K. Lin, et al., Second-order
topology and multi-dimensional topological transitions in sonic crystals,
Nat. Phys. 15 (2019) 582.

\bibitem{LLuPRL20} H. Cheng, Y. Sha, R. Liu, et al., Discovering topological
surface states of Dirac points, Phys. Rev. Lett. 124 (2020) 104301.

\bibitem{CQiuPRL22} J. Du, T. Li, X. Fan, et al., Acoustic realization of
surface-obstructed topological insulators, Phys. Rev. Lett. 128 (2022)
224301.

\bibitem{XDZhangPRL21} W. Zhang, D. Zou, Q. Pei, et al., Experimental
Observation of Higher-Order Topological Anderson Insulators, Phys. Rev.
Lett. 126 (2021) 146802.

\bibitem{GMaNRP19} G. Ma, M. Xiao, C. T. Chan, Topological phases in
acoustic and mechanical systems, Nat. Rev. Phys. 1 (2019) 281.
\end{thebibliography}
\end{document}


\section*{Supplementary Materials for \textquotedblleft Synchronized in-gap
edge states and robust copropagation in topological insulators without
magnetic flux"}

\begin{center}
Liangcai Xie, Tianyi He, and Liang Jin$^{\ast }$\\[2pt]
\textit{School of Physics, Nankai University, Tianjin 300071, China}
\end{center}

\subsection{The polarization for topological characterization}

The polarization of the square lattice in Fig.~3(a) of the main text is
visualized in Fig.~4(a2), Fig.~4(b2), and Fig.~4(c2) of the main text. In
the following, we prove the validity in detail. The eigenstate of the Bloch
Hamiltonian $h\left( \mathbf{k}\right) =\mathbf{d}\left( \mathbf{k}\right)
\cdot \mathbf{\sigma +}d_{0}\left( \mathbf{k}\right) \sigma _{0}$ can be
written in the form of%
\begin{equation}
\left\vert \psi _{\pm }\left( \mathbf{k}\right) \right\rangle =\frac{1}{%
\sqrt{2}}\left(
\begin{array}{c}
\pm 1 \\
e^{i\varphi \left( \mathbf{k}\right) }%
\end{array}%
\right) ,e^{i\varphi \left( \mathbf{k}\right) }=\frac{d_{x}+id_{y}}{%
\left\vert \mathbf{d}\right\vert }.  \label{ES}
\end{equation}%
The $d_{0}\left( \mathbf{k}\right) \sigma _{0}$ term in the Bloch
Hamiltonian does not alter the eigenstate. The polarization defined by the
eigenstate of $h\left( \mathbf{k}\right) $ is identical to the polarization
defined by the eigenstate of $\mathbf{d}\left( \mathbf{k}\right) \cdot
\mathbf{\sigma }$. The effective magnetic field term $\mathbf{d}\left(
\mathbf{k}\right) \cdot \mathbf{\sigma }=d_{x}\sigma _{x}+d_{y}\sigma _{y}$\
of the square lattice has the chiral symmetry. The polarization of a system
with chiral symmetry is associated with the winding of effective magnetic
field.

The components of the wave polarization projected in the $x$ and $y$
directions are defined as%
\begin{eqnarray}
P_{\pm ,x} &=&\frac{1}{\left( 2\pi \right) ^{2}}\int_{-\pi }^{\pi
}\int_{-\pi }^{\pi }-i\left\langle \psi _{\pm }(k_{x},k_{y})|\partial
_{k_{x}}\psi _{\pm }(k_{x},k_{y})\right\rangle dk_{x}dk_{y}  \label{Px} \\
P_{\pm ,y} &=&\frac{1}{\left( 2\pi \right) ^{2}}\int_{-\pi }^{\pi
}\int_{-\pi }^{\pi }-i\left\langle \psi _{\pm }(k_{x},k_{y})|\partial
_{k_{y}}\psi _{\pm }(k_{x},k_{y})\right\rangle dk_{x}dk_{y}  \label{Py}
\end{eqnarray}%
The Zak phase at fixed $k_{y}$ is defined as $Z_{\pm ,x}(k_{y})=\int_{-\pi
}^{\pi }-i\left\langle \psi _{\pm }(k_{x},k_{y})|\partial _{k_{x}}\psi _{\pm
}(k_{x},k_{y})\right\rangle dk_{x}$. From the eigenstate in Eq.~(\ref{ES}),
we obtain $Z_{\pm ,x}(k_{y})=\int_{-\pi }^{\pi }d\varphi \left( \mathbf{k}%
\right) /2$. Then, we obtain the polarization in the form of Zak phase as $%
P_{\pm ,x}=\left( 2\pi \right) ^{-2}\int_{-\pi }^{\pi }Z_{\pm
,x}(k_{y})dk_{y}$, i.e.,%
\begin{equation}
P_{\pm ,x}=\frac{1}{8\pi ^{2}}\int_{-\pi }^{\pi }d\varphi \left( \mathbf{k}%
\right) \int_{-\pi }^{\pi }dk_{y}.
\end{equation}%
The integral of the argument $\varphi \left( \mathbf{k}\right) $\ is
associated with the winding of the effective magnetic field $\mathbf{d}%
\left( \mathbf{k}\right) $. In the nontrivial phase with the in-gap edge
state, $\varphi \left( \mathbf{k}\right) $ varies $\pm 2\pi $ as $k_{x}$
varying an entire period from $-\pi $ to $\pi $ for every fixed $k_{y}$.
Thus, the Zak phase is $\pm \pi $ for every fixed $k_{y}$ and $P_{\pm
,x}=\pm 1/2$. We observe in Fig.~4(a2) of the main text that for any $k_{y}$%
, the effective magnetic field marked by the arrows rotates once as $k_{x}$
varying an entire period from $-\pi $ to $\pi $. The polarization in the $y$
direction has similar result as visualized in Fig.~4(b2).

\subsection{The absence of edge states in the inseparable region of
topological metallic phase}

The symmetry protected quantized polarization captures topologically in-gap
edge states. In the square lattice, although the chiral symmetry breaks, the
inversion symmetry is preserved and results in the quantized polarization.
The chiral symmetry breaking arises from the emergence of the identical
matrix $d_{0}\mathbf{\left( \mathbf{k}\right) }=2J\cos k_{y}$ in the Bloch
Hamiltonian, which can also alter the system from a topological insulator to
a topological metal. The topological metal does not have a full band gap.
The two energy bands of the one-dimensional projection of topological metal
may partially cover each other. Then, the edge states predicted by the
nonzero polarization disappear within the inseparable region of the energy
band.

\begin{figure}[tb]
\includegraphics[bb=10 0 565 230,width=18.8 cm]{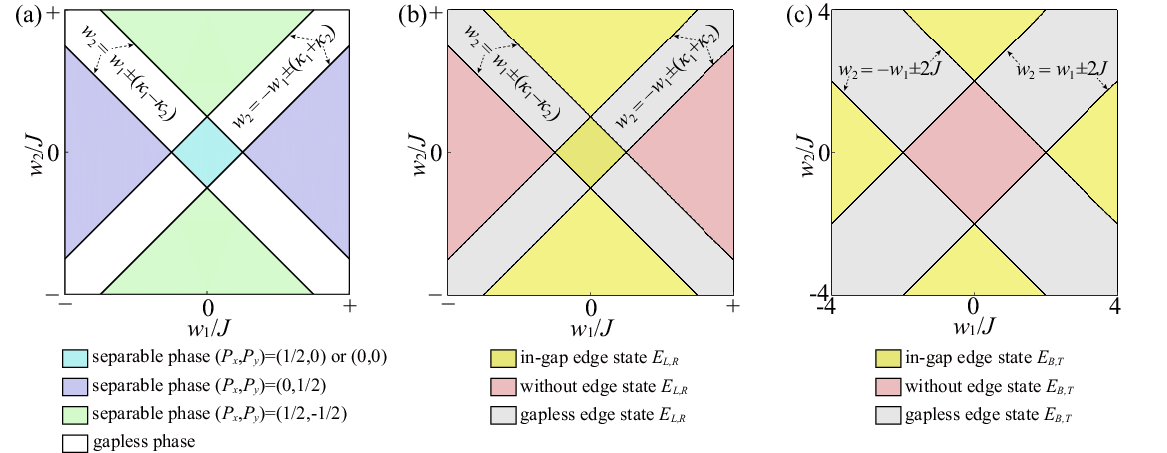}
\caption{(a) Polarization distribution. (b) In-gap state distributions under
the OBC in the $x$-direction. (c) In-gap state distributions under the OBC
in the $y$-direction. (a) and (b) are dependent of $\protect\kappa_{1}$ and $\protect\kappa_{2}$. (c) is independent of $\protect\kappa_{1}$ and $\protect\kappa_{2}$. Without the energy band overlap, the nonzero polarization
characterizes the presence of edge states.} \label{figS1}
\end{figure}

Since $d_{0}\mathbf{\left( \mathbf{k}\right) }=2J\cos k_{y}$ is a function
of the momentum $k_{y}$, thus, the one-dimensional projection energy band in
the $y$-direction is always separable.\ The in-gap edge states $%
E_{L,R}(k_{y})$\ predicted by the bulk polarization $P_{x}$\ are unaffected.
The absence of edge states in the inseparable region of the topological
metallic phase only occurs for the edge states $E_{B,T}(k_{x})$ in the other
direction.

The edge states are typically characterized by their localization, and in
the next section we analytically solve the edge states in both directions.
We take the left and right edge states as an example to determine the
presence of in-gap edge states. For the left and right edge states $%
E_{L,R}(k_{y})$, if the decay factor $|\rho _{L,R}|<1$ is satisfied for the
momentum $k_{y}\in \lbrack -\pi ,\pi ]$, the left and right edge states must
be in-gap throughout the Brillouin zone; if the decay factor $|\rho
_{L,R}|<1 $ is satisfied only for part of the momentum $k_{y}\in \lbrack
-\pi ,\pi ]$, then the edge states are truncated and there is an overlap of
the projected energy bands; if the decay factor $|\rho _{L,R}|<1$ is not
satisfied for any momentum $k_{y}\in \lbrack -\pi ,\pi ]$, then the edge
states are absent. The results obtained from this analysis are consistent
with the topological phase diagram, which is shown in Supplemental Figures %
\ref{figS1}(a) and \ref{figS1}(b). That is, the boundary determining the
in-gap edge state $E_{L,R}(k_{y})$ is the band gap closing boundary $%
w_{2}=w_{1}\pm \left( \kappa _{1}-\kappa _{2}\right) $ and $w_{2}=-w_{1}\pm
\left( \kappa _{1}+\kappa _{2}\right) $.

\begin{figure}[h]
\includegraphics[bb=0 0 505 310,width=18.8 cm]{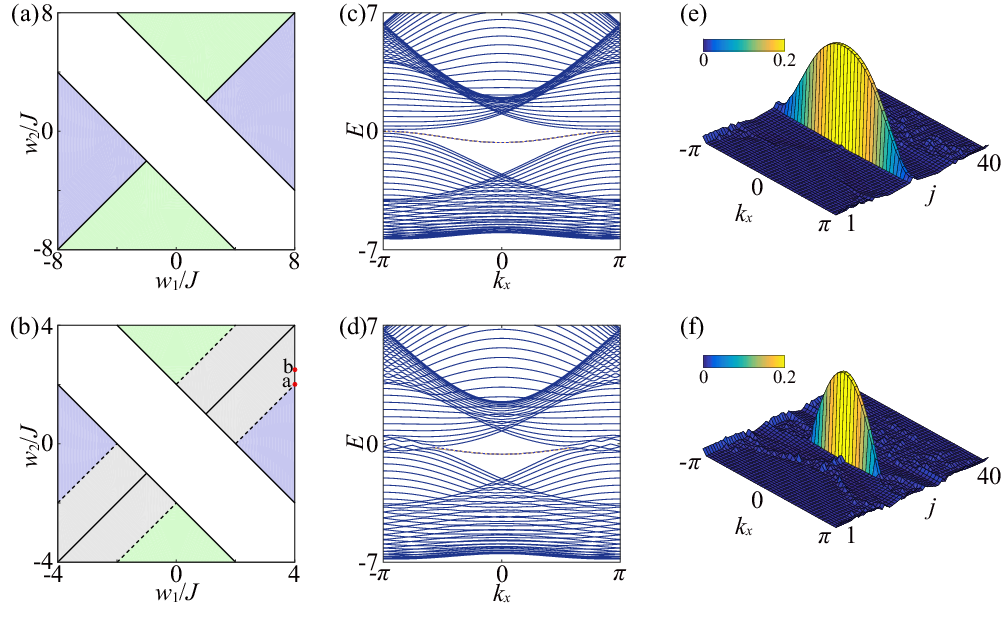}
\caption{(a) Topological phase diagrams not considering band overlap. (b)
Topological phase diagrams with band overlap. The parameters are set to $\protect\kappa _{1}=\protect\kappa _{2}=J$ for (a) and (b). (c) and (e)
In-gap edge states with no overlap and inverse participation ratio for all
states. (d) and (f) Gapless edge states with overlap and inverse
participation ratio for all states. The parameters are set to $w_{1}=4J$,
and $w_{2}=2J$ in (c) and (e) as indicated by the red dot a in (b) and $w_{2}=2.5J$ in (d) and (f) as indicated by the red dot b in (b).}
\label{figS2}
\end{figure}

The distribution boundaries for the bottom and top states $E_{B,T}\left(
k_{x}\right) $ are $w_{2}=\pm w_{1}\pm 2J$ as shown in Supplemental Figure %
\ref{figS1}(c). The yellow regions indicate the presence of in-gap edge
states without energy overlap in the projection bands, the gray regions
indicate the presence of gapless edge states with energy overlap in the
projection bands, and the red regions indicate the absence of in-gap edge
states. Obviously, these boundaries are different from the boundaries of
topological phase transition of Supplemental Figure \ref{figS1}(a). By
combining these two kinds of boundaries, we completely capture the in-gap
edge states. If $\left\vert \kappa _{1}+\kappa _{2}\right\vert \geq 2J$ and $%
\left\vert \kappa _{1}-\kappa _{2}\right\vert \geq 2J$, the energy overlap
is absent in the projection bands and the in-gap edge states $E_{B,T}(k_{x})$
predicted by $P_{y}$ are not affected. The energy overlap is present in the
projection bands for all the other cases. For example, when $\kappa _{1}=J$,
$\kappa _{2}=4J$, the projection bands do not have energy band overlap, and
the in-gap edge states are present in both the green and purple regions of
Fig.~3(c) of the main text. In this case, the bulk polarization $P_{y}$
accurately captures the in-gap edge states and the bulk-boundary
correspondence is valid in the entire Brillouin zone.

Supplemental Figure \ref{figS2} demonstrates a typical example with energy
overlap in the projection bands. When $\kappa _{1}=J$, $\kappa _{2}=J$, the
topological phase diagram without considering energy band overlap is shown
in Supplemental Figure \ref{figS2}(a). From the boundaries of the edge state
distribution, the energy band overlap is present in the projection bands,
and the in-gap edge states disappear in the gray regions as compared in
Supplemental Figure \ref{figS2}(b), i.e., the in-gap edge states predicted
by $P_{y}$ are destroyed by the energy overlap. The purple regions do not
have energy overlap in the projection bands, and the in-gap edge states
predicted by $P_{y}$ are not affected. The dashed black line $w_{2}=\pm
w_{1}\pm 2J$ divides the region with and without energy band overlap. The
white regions are the gapless phase with inseparable energy bands.

Supplemental Figures \ref{figS2}(c)-\ref{figS2}(f) demonstrate the energy
band overlap under two parameters as indicated by the red dots in
Supplemental Figure \ref{figS2}(b). Supplemental Figure \ref{figS2}(c)
illustrates the critical case that the projected energy bands under open
boundary condition in the $y$-direction are touched. The in-gap edge states $%
E_{B(T)}$ and the bulk bands intersect at $k_{x}=\pi $. Supplemental Figure %
\ref{figS2}(d) illustrates the case that the projected bands under open
boundary condition in the $y$-direction have overlap. The in-gap edge states
$E_{B(T)}$ are partially dissolved into the bulk bands, as evidenced from
the inverse participation ratio shown in Supplemental Figure \ref{figS2}(f).
Notably, the in-gap edge state is completely dissolved into the bulk in
certain cases of the gray regions.

\subsection{The edge states of square lattice}

The edge states with energies $E_{L,R}\left( k_{y}\right) $ exist under the
open boundary condition in the $x$ direction, and the edge states with
energies $E_{B,T}\left( k_{x}\right) $ exist under the open boundary
condition in the $y$\ direction. The wave functions for the edge states are
expressed as $|\psi _{L,R,B,T}\rangle =\left( \varphi _{1,A},\varphi
_{1,B},\cdots ,\varphi _{n,A},\varphi _{n,B},\cdots ,\varphi _{N,A},\varphi
_{N,B}\right) $, where $N$ is the total number of unit cells along the
direction with the open boundary condition and $n$ indexes the unit cell. $%
\varphi _{n,A}$ ($\varphi _{n,B}$) represents the component of wave function
on the sublattice $A$ ($B$) in the $n$-th unit cell. The wave functions of
the four edge states are analytically obtained at the infinite-size
limitation $N\rightarrow \infty $.

Applying the Fourier transformation in the $y$ direction, we obtain the
one-dimensional projection lattice as schematically illustrated in
Supplemental Figure \ref{figS3}(a). The one-dimensional projection lattice
is a Rice-Mele chain with a $k_{y}$-dependent uniform coupling along the $x$
direction and the sites of sublattices $A$ and $B$ alternately appear. The
wave functions $|\psi _{L}\rangle $ and $|\psi _{R}\rangle $ are for the
edge states with energies $E_{L}\left( k_{y}\right) $ and $E_{R}\left(
k_{y}\right) $
\begin{equation}
E_{L,R}\left( k_{y}\right) =2J\cos k_{y}
\end{equation}%
For the edge state $|\psi _{L}\rangle $, the component of $|\psi _{L}\rangle
$ at the sublattice $B$ vanishes ($\varphi _{n,B}=0$). For the edge state $%
|\psi _{R}\rangle $, the component of $|\psi _{R}\rangle $ at the sublattice
$A$ vanishes ($\varphi _{n,A}=0$). The wave functions are expressed as
\begin{equation}
|\psi _{L}\rangle =(1,0,\cdots ,\rho _{L}^{n-1},0,\cdots ,\rho
_{L}^{N-1},0),|\psi _{R}\rangle =(0,\rho _{R}^{N-1},\cdots ,0,\rho
_{R}^{N-n},\cdots ,0,1),
\end{equation}%
where the decay factors are $\rho _{L}=-\left( \kappa
_{1}+w_{1}e^{ik_{y}}\right) /\left( \kappa _{2}+w_{2}e^{-ik_{y}}\right) $
and $\rho _{R}=\rho _{L}^{\ast }$. As shown in Fig.~4 of the main text, the
edge state $|\psi _{L}\rangle $ only localizes on the sublattice $A$ and
decays exponentially from left to right of the square lattice; and the edge
state $|\psi _{R}\rangle $ only localizes on the sublattice $B$ and decays
exponentially from right to left of the square lattice.

\begin{figure}[tb]
\includegraphics[bb=20 0 470 85,width=16 cm]{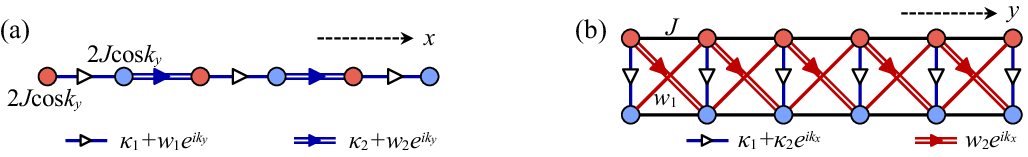}
\caption{Schematic of the 1D projection lattice for the 2D square lattice.
The (a) Rice-Mele chain and (b) Creutz ladder are obtained through applying
the Fourier transformation in the $y$ direction and in the $x$ direction,
respectively.} \label{figS3}
\end{figure}

Applying the Fourier transformation in the $x$ direction, we obtain the
one-dimensional projection lattice as schematically illustrated in
Supplemental Figure \ref{figS3}(b). The one-dimensional projection lattice
is a Creutz ladder with the two legs along the $y$ direction, one ladder-leg
site belongs to the sublattice $A$ and the other ladder-leg site belongs to
the sublattice $B$. The ladder rung coupling is $k_{x}$-dependent. The wave
functions $|\psi _{B}\rangle $ and $|\psi _{T}\rangle $ are for the edge
states with energies $E_{B}\left( k_{x}\right) $ and $E_{T}\left(
k_{x}\right) $. Solving the Hamiltonian, we observe that the energy of edge
state is determined from the following expression
\begin{equation}
\left( \beta w+w_{2}e^{-ik_{x}}\right) \left( E-\left( \kappa _{1}+\kappa
_{2}e^{ik_{x}}\right) \beta w\right) =\left( w_{1}\beta w+1\right) \left(
E\beta w-\left( \kappa _{1}+\kappa _{2}e^{-ik_{x}}\right) \right) .
\label{LR}
\end{equation}%
where $w=w_{1}-w_{2}e^{-ik_{x}}$. The Hermiticity of the system results in $%
E_{B,T}\left( k_{x}\right) =Re(E)$ and $\beta $\ is determined from $Im(E)=0$%
. We obtain $\beta $ in the form of%
\begin{equation}
\beta =\frac{-w_{1}^{2}+w_{2}^{2}-\sqrt{\left( w_{1}^{2}-w_{2}^{2}\right)
^{2}-4J^{2}\left( w_{1}^{2}+w_{2}^{2}\right) +8J^{2}w_{1}w_{2}\cos k_{x}}}{%
2J\left( w_{1}^{2}+w_{2}^{2}-2w_{1}w_{2}\cos k_{x}\right) }.
\end{equation}%
It is noted that $\beta $\ is independent of $\kappa _{1}$\ and $\kappa _{2}$%
. For the edge state $|\psi _{B}\rangle $, if we take $\varphi _{1,B}=1$,
the wave function of bottom state is expressed%
\begin{equation}
\left\{
\begin{array}{l}
\varphi _{2,B}=\frac{E\beta w-\left( \kappa _{1}+\kappa
_{2}e^{-ik_{x}}\right) }{J\beta w+w_{2}e^{-ik_{x}}}\varphi _{1,B}, \\
\varphi _{n,A}=\beta w\varphi _{n,B}, \\
\varphi _{n,B}=\frac{E\beta w-\left( \kappa _{1}+\kappa
_{2}e^{-ik_{x}}\right) }{J\beta w+w_{2}e^{-ik_{x}}}\varphi _{n-1,B}-\frac{%
J\beta w+w_{1}}{J\beta w+w_{2}e^{-ik_{x}}}\varphi _{n-2,B},\left( 3\leq
n\leq N\right) .%
\end{array}%
\right.
\end{equation}

For the edge state $|\psi _{T}\rangle $, if we take $\varphi _{N,A}=1$, the
wave function of top state is expressed as%
\begin{equation}
\left\{
\begin{array}{l}
\varphi _{N-1,A}=\frac{E\beta w^{\ast }-\left( \kappa _{1}+\kappa
_{2}e^{ik_{x}}\right) }{J\beta w^{\ast }+w_{2}e^{ik_{x}}}\varphi _{N,A}, \\
\varphi _{N-n+1,B}=\beta w^{\ast }\varphi _{N-n+1,A}, \\
\varphi _{N-n+1,A}=\frac{E\beta w^{\ast }-\left( \kappa _{1}+\kappa
_{2}e^{ik_{x}}\right) }{J\beta w^{\ast }+w_{2}e^{ik_{x}}}\varphi _{N-n+2,A}-%
\frac{J\beta w^{\ast }+w_{1}}{J\beta w+w_{2}e^{ik_{x}}}\varphi
_{N-n+3,A},\left( 3\leq n\leq N\right) .%
\end{array}%
\right.
\end{equation}

For topological nontrivial phases of $P_{y}=1/2$ with $|w_{1}|>|w_{2}|$, the
solution of $\beta $ satisfies $|\beta w|>1$. Consequently, the edge states $%
|\psi _{B}\rangle $ and $|\psi _{T}\rangle $\ are predominantly distributed
on the sublattices $A$ and $B$, respectively. Conversely, for topological
nontrivial phases of $P_{y}=-1/2$\ with $|w_{1}|<|w_{2}|$, the situation is
reversed, with the edge states $|\psi _{B}\rangle $ and $|\psi _{T}\rangle $
predominantly distributed on the sublattices $B$ and $A$, respectively. The
edge state $|\psi _{B}\rangle $ decays exponentially from bottom to top of
the square lattice, while the edge state $|\psi _{T}\rangle $ decays
exponentially from top to bottom of the square lattice as illustrated in
Fig.~4 of the main text.

\subsection{Different topological phases of the square lattice model and
two-dimensional Su-Schrieffer-Heeger model}

We emphasize that the square lattice model supporting the first-order
topological phase is intrinsically different from these two-dimensional SSH
models supporting the higher-order topological phase in terms of symmetry,
topological classification, edge states, and corner states. (i) The chiral
symmetry is broken in the square lattice model, whereas it preserves in the
two-dimensional SSH model. Then, the former belongs to the AI class, whereas
the latter belongs to the BDI class in the tenfold-way topological
classification. (ii) The edge states of the square lattice model are a pair
of edge states in each direction, whereas the edge states of the
two-dimensional SSH model are two pairs of edge states as a consequence that
the former is a two-band model and the latter is a four-band model. (iii)
The quantized bulk polarization $P_{x}\cdot P_{y}\neq 0$ $\left( =0\right) $
in the two-dimensional SSH model indicates the presence (absence) of corner
states, whereas this is not necessarily valid in the square lattice model.

\begin{figure}[t]
\includegraphics[bb=0 0 250 240,width=12.8 cm]{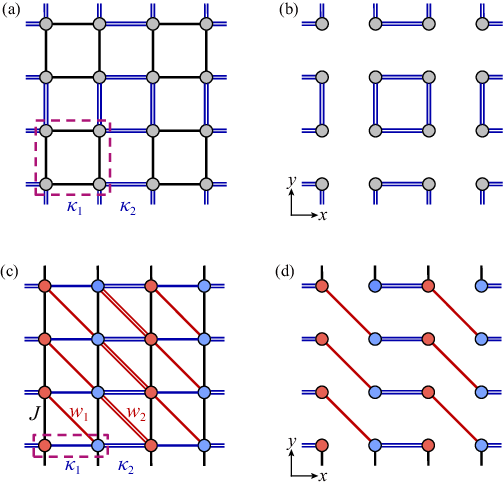}
\caption{(a) Schematic of the two-dimensional SSH model. (b) Four corner
states of (a) at $\protect\kappa_1=0$, $\protect\kappa_2\neq0$. (c)
Schematic of the square lattice model. (d) Two corner modes appear at the
lower-left and upper-right corners of (c) at $\protect\kappa_1=0$, $\protect\kappa_2\neq0$, $w_1\neq0$, $w_2=0$, $J=0$.}
\label{figS4}
\end{figure}

The two-dimensional Su-Schrieffer-Heeger (SSH) model [Supplemental Figure~%
\ref{figS4}(a)] generally supports two types of higher-order topological
phases with topological corner states, one type with quantized quadrupole
moment and the other type without quadrupole moment. The former case is the
Benalcazar-Bernevig-Hughes (BBH) model, which is a two-dimensional SSH model
with $\pi $\ flux in each plaquette and the two-fold degenerate energy bands
\cite{HughesScience17}. The quadrupole moments are associated with edge
polarizations and are protected by $C_{\mathbf{4}}$\ symmetry, which
guarantees the quantized corner charge at each corner, manifesting as the
existence of four degenerate corner states respectively localized at the
four corners [Supplemental Figure \ref{figS4}(b)]. The latter case is a
two-dimensional SSH model without any flux in each plaquette and the single
occupied energy band. The quadrupole moment is absent. The edge
polarizations reduce to the bulk polarizations and the existence of corner
state is determined by the nonvanishing values of $P_{x}\cdot P_{y}$\ \cite%
{FLiuPRL19}. Then, four corner states appear only if both $P_{x}$\ and $%
P_{y} $\ are nonzero. The two-dimensional SSH model also supports
copropagation in the higher-order topological phase with $\left(
P_{x},P_{y}\right) =\left( 1/2,1/2\right) $. The spectrum for the
two-dimensional SSH model under open boundary condition exhibits two pairs
of in-gap edge states in both directions. The edge state excitations within
the band gap copropagate along the boundaries and are reflected back at the
corners because of the mismatch of the on-resonant state in the other
direction.

The square lattice model [Supplemental Figure \ref{figS4}(c)] has the
topological phase $\left( P_{x},P_{y}\right) =\left( 1/2,-1/2\right) $.
Since the $C_{2}$\ symmetry of the system is preserved while the $C_{4}$
symmetry is broken, the polarization is quantized, but the corner states
are\ not topologically protected. The corner states still exist even if $%
P_{x}\cdot P_{y}=0$. For instance, at $J=\kappa _{1}=w_{2}=0$, if $\kappa
_{2}<w_{1}$, the polarizations are $P_{x}=0$ and $P_{y}=1/2$. Under the open
boundary conditions in both horizontal and vertical directions, the single
sites at the lower-left and upper-right corners are decoupled from the other
sites, forming two corner states [Supplemental Figure \ref{figS4}(d)]. In
addition, the nearest-neighbor coupling $J$ may also affect the existence of
corner states.

{\color{red} }

\subsection{Copropagation of in-gap edge states in the square lattice}

\begin{figure}[thb]
\includegraphics[bb=0 0 420 180,width=18.3cm]{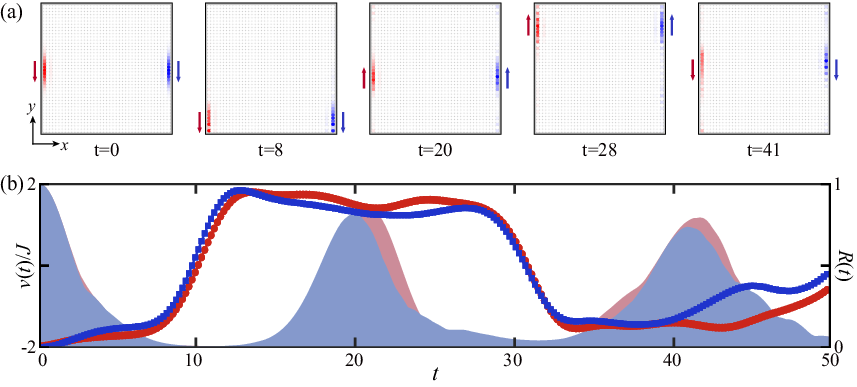}
\caption{(a) Snapshots of copropagation along the vertical
boundaries and (b) the propagation velocity and return probability for the
couplings randomly deviated from the set strengths within the range of $[-10\%,10\%]$. The Gaussian wave packet of the initial edge state excitation on both the left and right boundaries has the momentum $k_{y}=\protect\pi/2$. Red and blue lines indicate the propagation velocities $\protect\nu (t)$ of the wave packets on the left and right boundaries, respectively. Red and blue areas
indicate the return probabilities $R(t)$ of the wave packets on the left and right boundaries, respectively. The lattice size is $40\times 40$, the system parameters are from Fig.~4(a).
The unit of time is $J^{-1}$.} \label{figS6}
\end{figure}

In the topological insulating phase of the time-reversal symmetric square
lattice, the bulk states are completely off-resonant with the in-gap edge
states. Thus, the edge state excitations propagating along the boundaries
are reflected back toward the opposite direction at the corners. Similarly,
the large obstacles laid on the boundaries of the square lattice may cause
backscattering of the edge state excitations. The presence of backscattering
channels is the reason for the fact that the disorder in principle must not
be too large. However, the presence of backscattering channels does not mean
that the dynamics of edge state excitations along the boundaries are not
robust to disorder. The band gap can protect the robust propagation along
the boundaries.

In Fig.~5 of the main text, we show the robust copropagation of the
synchronized in-gap edge states for the weak disorder within the range of $%
[-5\%,5\%]$. Supplemental Figure \ref{figS6} shows the robust copropagation
of the synchronized in-gap edge state for the moderate disorder within the
range of $[-10\%,10\%]$. From the numerical simulation, we notice that the
disorder does not obviously affect the robust copropagations of the
synchronized in-gap edge states even if the deviation from the exact values
of the couplings reaches the range of $[-10\%,10\%]$. The backscattering is
observed in the copropagation of the synchronized in-gap edge state for the
large disorder within the range of $[-15\%,15\%]$.

{\color{red} }
\subsection{Counterpropagation of in-gap edge states in the square lattice}

\begin{figure}[h]
\includegraphics[bb=0 0 420 180,width=18.3cm]{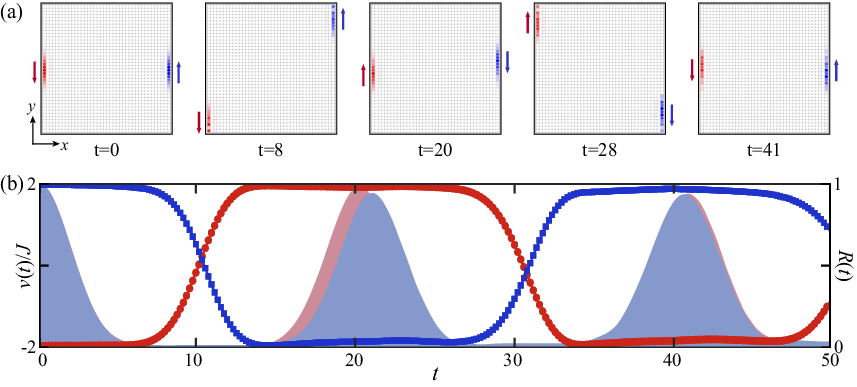}
\caption{(a) Snapshots of counterpropagation along the vertical
boundaries and (b) the propagation velocity and return probability for the
couplings randomly deviated from the set strengths within the range of $[-5\%,5\%]$.
The Gaussian wave packet of the initial edge state excitation on the left boundary has the momentum $k_{y}=\protect\pi/2$,
and the Gaussian wave packet of the initial edge state excitation on the right boundary has the momentum $k_{y}=-\protect\pi/2$.
Red and blue lines indicate the propagation velocities $\protect\nu (t)$ of the wave packets on the left and right boundaries, respectively. Red and blue areas
indicate the return probabilities $R(t)$ of the wave packets on the left and right boundaries, respectively. The lattice size is $40\times 40$, the system parameters are from Fig.~4(a).
The unit of time is $J^{-1}$.} \label{figS5}
\end{figure}

The bidirectionality is a key feature of the in-gap edge states in the
time-reversal symmetric topological insulators. Consequently, this naturally
provides a possibility for picking the propagation direction along the
boundaries in a desirable manner through selectively exciting the edge
states. The copropagation of in-gap edge states along the parallel
boundaries is realized when the excitation of in-gap edge states on the
parallel boundaries has the same momentum. By contrast, the
counterpropagation of in-gap edge states along the parallel boundaries is
realized when the excitation of in-gap edge states on the parallel
boundaries has the opposite momenta.

Supplemental Figure \ref{figS5} shows the numerical simulation of the robust
counterpropagation of the in-gap edge state excitations. The momentum for
the edge state excitation on the left boundary is $k_{y}=\pi /2$\ and the
momentum for the edge state excitation on the right boundary is $k_{y}=-\pi
/2$. The parameters of the square lattice are chosen as the topological
phase $(P_{x},P_{y})=(1/2,0)$ in Fig.~4(a) of the main text. The coupling
disorder is set within the range of $[-5\%,5\%]$. Notably, the
counterpropagations along both the left and right boundaries are not
obviously affected by the disorder.

\vspace{1.0cm} $^*$ jinliang@nankai.edu.cn